\documentclass[12pt]{iopart}

\usepackage{bm}
\usepackage{epsfig}
\usepackage{citesort}
\usepackage{graphicx}
\eqnobysec

\begin{document}


\begin{flushright}
{\large \tt LAPTH-1229/07}
\end{flushright}

\title[Using BBN in cosmological parameter extraction from CMB]{Using
BBN in cosmological parameter extraction from CMB:
a forecast for {\sc PLANCK}}

\author{Jan~Hamann$^1$, Julien Lesgourgues$^1$ and
Gianpiero Mangano$^{2}$}

\address{$^1$~LAPTH (Laboratoire d'Annecy-le-Vieux de Physique
Th\'eorique, CNRS UMR5108 \& Universit\'e de Savoie), BP 110, F-74941
 Annecy-le-Vieux Cedex, France\\ $^2$~Istituto Nazionale di Fisica
 Nucleare - Sezione di Napoli - \\ Complesso Universitario di Monte
 S.~Angelo, I-80126 Napoli, Italy}

\ead{\mailto{jan.hamann@lapp.in2p3.fr},
     \mailto{julien.lesgourgues@lapp.in2p3.fr},
     \mailto{mangano@na.infn.it}}

\begin{abstract}
  Data from future high-precision Cosmic Microwave Background (CMB)
  measurements will be sensitive to the primordial Helium abundance
  $Y_p$. At the same time, this parameter can be predicted from Big
  Bang Nucleosynthesis (BBN) as a function of the baryon and
  radiation densities, as well as a neutrino chemical potential. We
  suggest to use this information to impose a self-consistent BBN prior
  on $Y_p$ and determine its impact on parameter inference from
  simulated {\sc planck} data. We find that this approach can
  significantly improve bounds on cosmological parameters compared to
  an analysis which treats $Y_p$ as a free parameter, if the neutrino
  chemical potential is taken to vanish. We demonstrate that fixing
  the Helium fraction to an arbitrary value can seriously bias
  parameter estimates. Under the assumption of degenerate BBN (i.e.,
  letting the neutrino chemical potential $\xi$ vary), the BBN prior's
  constraining power is somewhat weakened, but nevertheless allows us
  to constrain $\xi$ with an accuracy that rivals bounds inferred from
  present data on light element abundances.
\end{abstract}
\maketitle

\section{Introduction}\label{sec:introduction}

Forthcoming experiments on Cosmic Microwave Background (CMB)
anisotropies such as {\sc planck}\footnote{ESA home page for the {\sc
planck} project: {\tt
http://astro.estec.esa.nl/SA-general/Projects/Planck/} and Planck-HFI
web site: {\tt http://www.planck.fr/}} \cite{planck}, combined with
other astrophysical observations, are expected to provide detailed
information on the cosmological model which describes the evolution of
the Universe. Their data will allow us to constrain the parameters of
these models with an unprecedented accuracy. In fact, to fully extract
information from such precise experimental data, it is crucial to have
detailed theoretical models to compare with, possibly using reliable
``priors'' on cosmological parameters or models which can be obtained
by independent theoretical tools or experimental data. This helps in
reducing the effect of parameter degeneracies which typically
limits the amount of information which can be obtained from data.

In this paper we present one example of this kind, by considering the
impact of a detailed estimate of the $^4$He mass fraction\footnote{We
  point out that as usual in the literature we refer to $Y_p$ as the
  ``$^4$He mass fraction'', though strictly speaking it is not correct
  at the percent level, for it neglects the contribution of nuclear
  binding energy.} $Y_p = 4 n_{\rm He}/n_{b}$ on CMB data analyses,
obtained from an accurate prediction of its value from Big Bang
Nucleosynthesis (BBN). In this framework, $Y_p$ is given as a function
of the baryon density and, in more exotic scenarios, extra
relativistic degrees of freedom and/or non-zero neutrino chemical
potentials $\mu_{\nu}$.

The amount of $^4$He nuclei produced during BBN plays a relevant
r\^ole at the epoch of recombination, as it is one of the parameters
controlling the evolution of the free electron fraction
\cite{Hu:1995fqa,Switzer:2007sq} (it also affects the later phase of
reionisation of the Universe, but this effect is extremely small
\cite{Lewis:2006ym}). Thus, the CMB power spectrum which is observed
today depends significantly on this parameter. Typically, in current
CMB analyses (and in a number of forecasts), this parameter is fixed
to a reference value $Y_p = 0.24$, suggested by independent
measurements obtained by studying extragalactic HII regions in blue
compact galaxies, which are however affected by quite a large
systematic uncertainty \cite{olive,Izotov:2007ed}. This approach is
satisfactory with present CMB data, but it is not fully correct as it
does not take into account the fact that indeed, $Y_p$ is strongly
correlated to other cosmological parameters, in particular the baryon
fraction $\Omega_{\rm b} h^2$ and the energy density $\rho_R$ in the
form of relativistic species, which is usually parameterised in terms
of the effective neutrino number $N_{\rm eff}= 3 + \Delta N$ such that
\begin{equation}
\rho_R = \frac{\pi^2}{15} T_\gamma^4  \left[ 1 + \frac{7}{8}
\left(\frac{4}{11} \right)^{4/3} (3+ \Delta N) \right],
\end{equation}
with $T_\gamma$ the photon temperature. The standard scenario of three
non-degenerate active neutrinos corresponds to $\Delta N= 0.046$ due to the
effect of non-thermal corrections to neutrino distributions during the
$e^+-e^-$ annihilation phase \cite{pinto}.


A different (and more consistent) approach is taken in
\cite{Trotta:2003xg}, where $Y_p$ is considered as a free parameter to
be obtained as a result of a likelihood analysis along with the other
relevant cosmological parameters, or rather fixed up to some given
error by adopting a gaussian prior. This approach leads to a more
correct estimate of the baryon density and, as we will also show in
the following, avoids biasing the point estimates of some cosmological
parameters, such as the scalar spectral index $n_s$.

In the present work we will consider yet a different approach
previously suggested by Ichikawa and Takahashi~\cite{Ichikawa:2006dt},
which explicitly takes into account the dependence of $Y_p$ on
$\Omega_{\rm b} h^2$ (and possibly also on $\Delta N$ and on the
neutrino chemical potentials) as obtained from BBN. We use the
scale-independent ratio $\xi \equiv \mu_{\nu}/ k T_{\nu}$, where
$T_\nu$ is the neutrino temperature and $\mu_\nu$ is the chemical
potential, which is assumed to be the same for all three active
neutrino species, due to the effect of flavour oscillations
\cite{Dolgov:2002ab,Wong:2002fa}.
In particular, the Helium mass fraction is considered as a known
function $Y_p^{\rm BBN}(\Omega_{\rm b} h^2$, $\Delta N$, $\xi)$ and is
thus neither fixed a priori, nor left as a free independent
parameter. This function can be obtained in the framework of standard
BBN theory and is found to be a smooth monotonically increasing
function of $\Omega_{\rm b} h^2$ and $\Delta N$, and decreasing with
$\xi$. In particular, the result adopted in the present paper has been
computed by using the public BBN code
\texttt{PArthENoPE} \cite{Pisanti:2007hk}.  This tool provides a
careful determination of light nuclei abundances, with very small
uncertainties. \texttt{PArthENoPE} is the result of a large
reanalysis, updating the whole nuclear reaction network, and also
including the effect of radiative corrections to neutron/proton weak
processes, and the detailed treatment of neutrino decoupling described
in \cite{pinto}. For a comprehensive discussions of these issues, see
e.g.~\cite{serpico}. The theoretical error on $Y_p^{\rm BBN}$ is of
the order of $0.0002$, thus at the level of per mille, over the
relevant ranges of ($\Omega_{\rm b} h^2$, $\Delta N$,
$\xi$)~\cite{serpico}. The accuracy is ultimately limited by the
present experimental uncertainty on the neutron lifetime. This tiny
error essentially amounts to imposing a consistency relation between
$Y_p$, $\Omega_{\rm b}h^2$, $\Delta N$ and $\xi$.

The goal of this paper is therefore to determine how the
assumption of a well motivated BBN prior on $Y_p$ will affect the
estimates of cosmological parameters from {\sc planck} data.  Of
course, the validity of this method is based on the assumption of
standard BBN (or degenerate BBN (dBBN) in section \ref{sec:dbbn},
where $\xi \neq 0$ will be considered). It also implicitly assumes
that the value of $N_{\rm eff}$ does not change from the BBN epoch
until last scattering. However, possible future evidence for
non-standard BBN scenarios or a more involved evolution of
relativistic degrees of freedom could similarly be accounted for
by using a different -- yet still theoretically calculable --
functional form of $Y_p$.

\section{Forecast for {\sc planck} data \label{sec:forecast}}

In our analysis, we compare three different ways of treating the
Helium mass fraction:

\begin{itemize}

    \item{Fixed $Y_p$:\\ As in most recent analyses and forecasts of
CMB data, $Y_p$ is fixed to a value of $Y_p = 0.24$. This is also
the default value set in \texttt{CosmoMC}/\texttt{CAMB}
\cite{Lewis:2002ah,Lewis:1999bs}. Note that if we set $N_{\rm
eff}$ to its standard value of 3.046 and assume no neutrino
asymmetry, the value of 0.24 is not consistent with current bounds
on the baryon density. At the moment, CMB data are not very
sensitive to $Y_p$, so the bias expected from fixing $Y_p$ in this
way is negligible.}

    \item{Free $Y_p$:\\
Here, no additional assumptions are made and the Helium mass
fraction is kept completely free, with a top hat prior ranging
from 0 to 1.}


    \item{BBN prior on $Y_p$:\\
Under the assumption that BBN proceeded in the standard way, and
that $N_{\rm eff}$ remains constant between BBN and last
scattering, we exploit the fact that $Y_p$ is related to
$\Omega_{\rm b} h^2$, $\Delta N$, and possibly $\xi$ if
neutrinos have a sizable chemical potential. In a first-order
approach, one could fix $Y_p$ to the BBN prediction $Y_p^{\rm
BBN}(\Omega_{\rm b}h^2, \Delta N, \xi)$ calculated by
\texttt{PArthENoPE}. That way, however, one would not take into
account the theoretical uncertainty in $Y_p^{\rm BBN}$. In order
to treat the uncertainty properly, we keep $Y_p$ a free parameter,
but, taking $\delta Y_p^{\rm BBN}$ to be gaussian, add the
following term to the negative logarithm of the likelihood
$\mathcal{L}$ of each point in parameter space:
\begin{equation}
\Delta(-\ln \mathcal{L}) =  \frac{1}{2} \; \left(
\frac{Y_p-Y_p^{\rm BBN}(\Omega_{\rm b}h^2, \Delta N,
\xi)}{\sigma(Y_p)} \right)^2.
\end{equation}
Since recomputing $Y_p^{\rm BBN}$ for each point would not be
practical, we interpolate its value from a pre-computed grid. To
account for errors introduced due to interpolation, we increase the
absolute error on $Y_p^{\rm BBN}$ to $\sigma(Y_p)=0.0003$. Note that we do not
employ any additional data here e.g., astrophysical measurements
of primordial element abundances, that may be subject to large
systematic errors.}

\end{itemize}

\subsection{Fiducial data and parameter inference}

Following the method described in detail in
\cite{Perotto:2006rj,Hamann:2007sk}, one can generate a set of mock
CMB data, using the projected specifications of the {\sc planck}
satellite \cite{planck} (see table~\ref{table:planck}). However, for
the purpose of forecasting errors, it is sufficient to replace the
power spectrum of the mock data by that of the fiducial model, which
stands for an average over many possible mock data
sets~\cite{Perotto:2006rj}.  The data set comprises the $TT$- and
$EE$-auto-correlation spectra as well as the $TE$-cross-correlation
spectrum for multipoles up to $\ell = 2500$, and we assume a sky
coverage of $f_{\rm sky} = 0.65$. In our fiducial model we impose the
standard BBN consistency relation, spatial flatness, and ignore tensor
modes; its parameter values are summarized in
table~\ref{table:params}.

\begin{table}[b!]
\caption{\label{table:planck} List of the experimental parameters
assumed for the {\sc planck} satellite \cite{planck}: $\theta_{\rm
beam}$ measures the width of the beam, $\Delta_{T,P}$ are the
sensitivities per pixel and $\nu$ is the center frequency of the
three HFI channels least affected by foregrounds.}\vskip5mm
\hskip45mm \footnotesize{
\begin{tabular}{cccc}
\br
 $\nu$/GHz & $\theta_{\rm beam}$ & $\Delta_T$/$\mu$K & $\Delta_P$/$\mu$K\\
\mr
 100 & 9.5' & 6.8 & 10.9\\
 143 & 7.1' & 6.0 & 11.4\\
 217 & 5.0' & 13.1 & 26.7\\
\br
\end{tabular}}
\end{table}

\begin{table}[t]
\caption{In this table we show the free parameters of our model,
their fiducial values used to generate the data set and the prior
ranges adopted in the analysis.\label{table:params}}\vskip5mm
\hskip15mm \footnotesize{\begin{tabular}{llll}
 \br
 Parameter&&Fiducial Value&Prior Range\\
 \mr
 Dark matter density & $\Omega_{\rm dm} h^2$ & 0.11 & $0.01\to0.99$ \\
 Baryon density & $\Omega_{\rm b} h^2$ & 0.022  & $0.005 \to 0.1$ \\
 Hubble parameter & $h$ & 0.7 & $0.4\to1$ \\
 Redshift of reionisation & $z_\mathrm{re}$  & 12 & $3\to50$\\
 Normalisation @ $k=0.002\mathrm{\ Mpc}^{-1}$&
 $\ln[10^{10}A_\mathrm{S}]$ & 3.264 & $2.7\to4$\\
 Scalar spectral index & $n_\mathrm{S}$ & 0.96  & $0.5\to1.5$\\
 Helium fraction & $Y_p$ & 0.2477 & $0\to1$\\
 Neutrino mass fraction & $f_\nu$ & 0 & $0\to1$\\
 Number of extra rel.~d.o.f. & $\Delta N$ & 0.046 & $-3\to4$\\
 Neutrino chemical potential & $\xi$ & 0 & $-1 \to 1$ \\
 \br
\end{tabular}}
\end{table}

We then perform the exercise of Bayesian parameter inference for a
number of models, differing in the number of basic free parameters
and the treatment of the primordial Helium fraction $Y_p$, as
described above. We first work under the standard assumption that
the chemical potential of neutrinos is negligible ($\mu_{\nu} \ll
k T_{\nu}$), and consider two basic models:

\begin{itemize}
    \item{A minimal model, with six free parameters ($\Omega_{\rm
b}h^2$, $\Omega_{\rm dm}h^2$, $H_0$, $z_{\rm re}$,
$\ln[10^{10}A_\mathrm{S}]$, $n_\mathrm{S}$), inspired by the current
``vanilla'' model. }
    \item{An extended model, where in addition to the parameters of
the minimal model we also vary the neutrino mass fraction
$f_\nu$ and the number of extra relativistic
degrees of freedom $\Delta N$. The introduction of $f_\nu$ is
motivated by the observation of neutrino oscillations, implying a
non-negligible effect of neutrino masses on cosmological perturbations
\cite{Lesgourgues:2006nd}. The neutrino fraction and $\Delta N$ are
known to be correlated in the analysis of CMB
data~\cite{Hannestad:2003ye,Crotty:2004gm,Hannestad:2006mi}.}
\end{itemize}

In section \ref{sec:dbbn}, we will repeat the analysis with one
extra free parameter which is known to modify the outcome of BBN
predictions: a non-zero chemical potential for neutrinos.

We employ a modified version of the Markov-Chain-Monte-Carlo code
\texttt{CosmoMC} \cite{Lewis:2002ah} to infer the posterior
probability density from the data. Eight Markov chains are generated
in parallel; their convergence is monitored with the help of the
Gelman-Rubin $R$-statistic \cite{gelru}, and our convergence criterion
is $R - 1 \leq 0.02$.

When facing real data, one could be worried that the theoretical
prediction for the anisotropy spectra might be insufficient. In
particular, issues like recombination or foreground contamination
need to be better understood. Since we use the same numerical code for
generating and analysing the data, we implicitly assume in this
forecast that all systematics are perfectly under control. Thus, our
inferred parameter errors may be slightly optimistic.

\begin{table}[t]
\caption{\label{table:sbbnerrors} This table shows the projected
absolute errors on the parameters of the two models with zero
neutrino chemical potential. For all parameters except the
neutrino fraction we quote the half width of the minimal 68\%
credible interval \cite{Hamann:2007pi}, for $f_\nu$ we give the
values of the 68\% upper limit (the lower limit being zero). The
columns labelled ``free'' show the results when leaving the Helium
fraction a free parameter, while those labelled ``BBN'' correspond
to the results imposing our BBN prior. Our results are in very
good agreement with those found in
\cite{Ichikawa:2006dt}.}\vskip5mm \hskip35mm
\footnotesize{\begin{tabular}{lllll} \br
    & \multicolumn{2}{c}{Minimal model} & \multicolumn{2}{c}{Extended model}\\ &
    \multicolumn{1}{c}{free} & \multicolumn{1}{c}{BBN} &
    \multicolumn{1}{c}{free} & \multicolumn{1}{c}{BBN}\\ \mr
    $\Omega_{\rm b} h^2$ & $2.2 \times 10^{-4}$ & $1.3 \times 10^{-4}$
    & $2.3 \times 10^{-4}$ & $2.2 \times 10^{-4}$\\ $\Omega_{\rm dm}
    h^2$ & $1.4 \times 10^{-3}$ & $1.2 \times 10^{-3}$ & $4.9 \times
    10^{-3}$ & $3.1 \times 10^{-3}$ \\ $h$ & $7.9 \times 10^{-3}$ &
    $5.9 \times 10^{-3}$ & $28 \times 10^{-3}$ & $24 \times 10^{-3}$
    \\ $z_\mathrm{re}$ & 0.40 & 0.39 & 0.41 & 0.41\\
    $\ln[10^{10}A_\mathrm{S}]$ & 0.024 & 0.015 & 0.024 & 0.021\\
    $n_\mathrm{S}$ & $7.2 \times 10^{-3}$ & $3.4 \times
    10^{-3}$ & $8.2 \times 10^{-3}$ & $7.7 \times 10^{-3}$ \\
    $Y_p$ & 0.011 & $3.1 \times 10^{-4}$ & 0.015 & $2.1
    \times 10^{-3}$ \\ $\Delta N$ & -- & -- & 0.26 & 0.15 \\
    $f_\nu$ & -- & -- & $\leq 0.041$ & $\leq 0.035$ \\ \br
\end{tabular}}
\end{table}

\subsection{Standard BBN}
\subsubsection{Minimal model}

\begin{figure}
\begin{center}
    \includegraphics[height=\textwidth,angle=270]{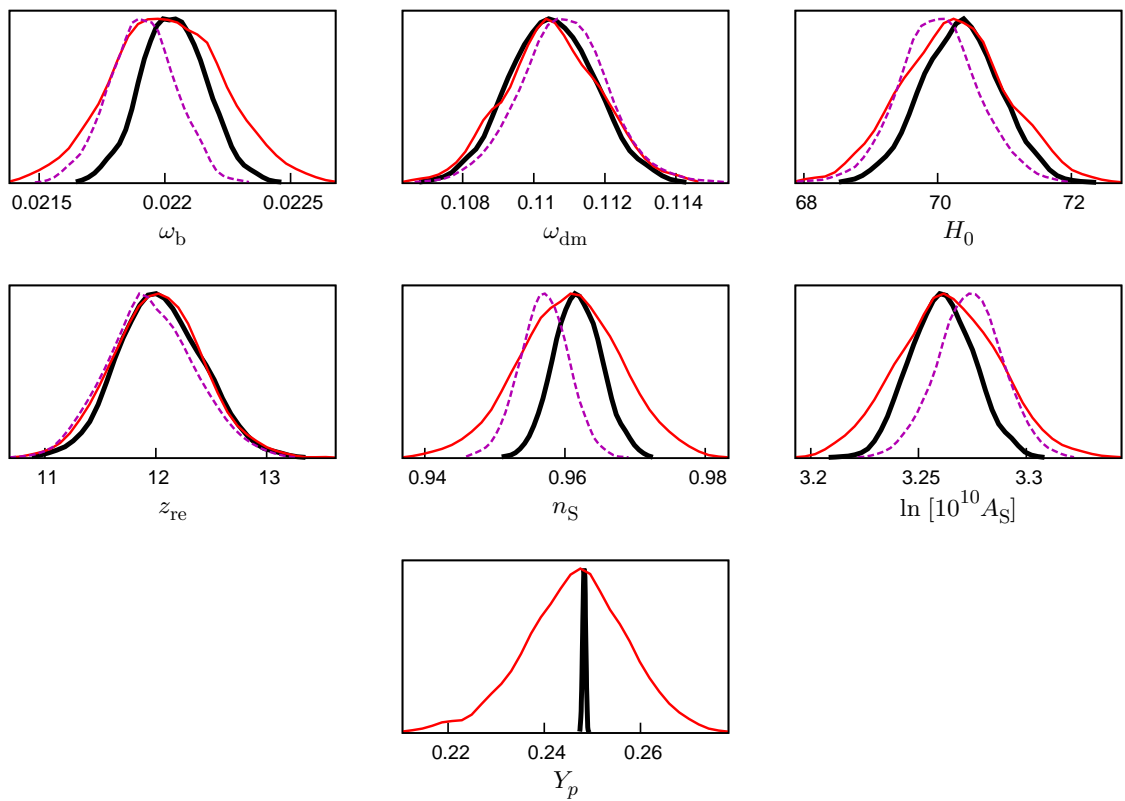}
\end{center}
\caption{\label{fig:7p1d} Marginalised posterior probabilities for the
parameters of the minimal model. The dashed purple curves correspond to
the case where $Y_p$ is fixed to an ``incorrect'' value of 0.24, the
red curves have $Y_p$ as a free parameter and the thick black curves
correspond to the case with a standard BBN prior.}
\end{figure}

The one-dimensional marginalised posterior probabilities for the
parameters of the minimal model are presented in figure
\ref{fig:7p1d}. A first striking observation is that fixing $Y_p$
``incorrectly'' to 0.24 leads to a significant bias of up to one
standard deviation in the point estimates for the baryon density,
Hubble parameter, spectral index and primordial spectrum
normalisation.  This is particularly worrisome for $n_{\rm S}$, since
bounds on this parameter are often used to constrain inflationary
models. We therefore strongly recommend not to fix $Y_p$ to some
arbitrary value, such as 0.24, in any analysis of future data.

\begin{figure}
\begin{center}
    \includegraphics[height=\textwidth,angle=270]{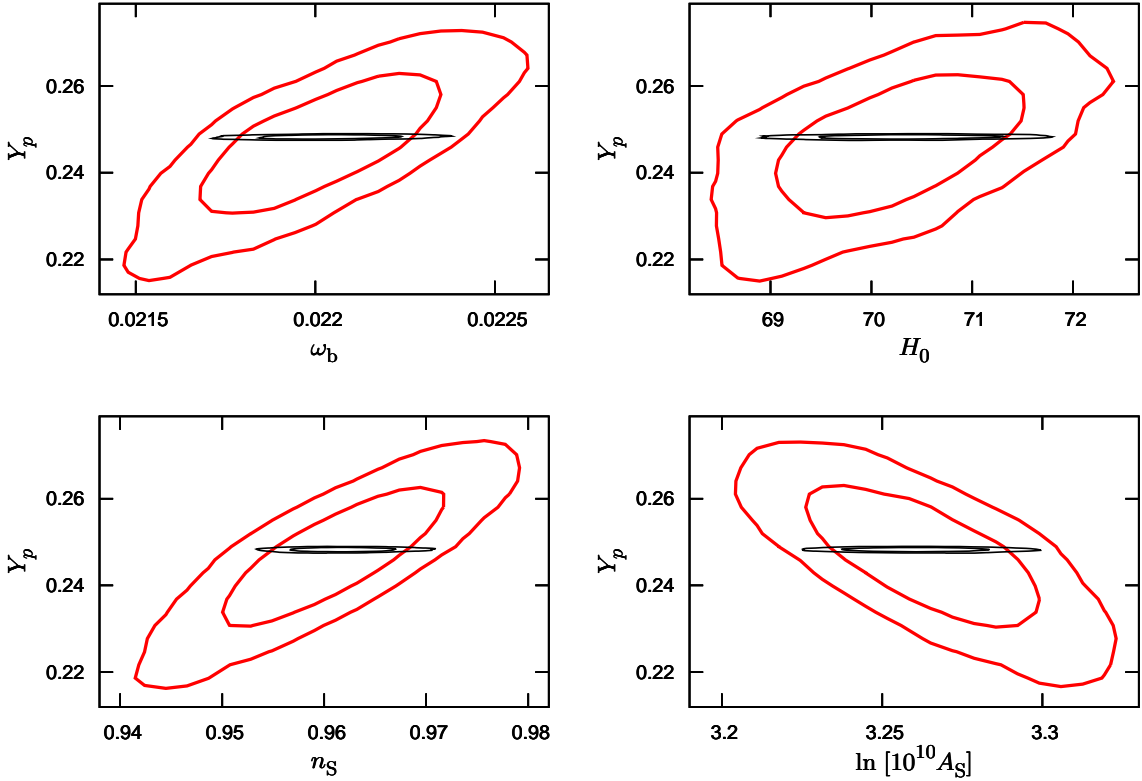}
\end{center}
\caption{\label{fig:7p2d} Two-dimensional marginalised joint
posterior 68\%- and 95\%-credible contours for the minimal model.
The thick red contours correspond to results with free $Y_p$, the
thin black contours represent the result when standard BBN is
imposed. This plot illustrates the degeneracies of $Y_p$ with
other cosmological parameters and shows how they can be broken by
imposing the BBN prior.}
\end{figure}

The reason for the bias in these parameters are degeneracies with the
Helium mass fraction; we illustrate these degeneracies in figure
\ref{fig:7p2d}. Of these four degeneracies, only the one with
$\omega_{\rm b}=\Omega_{\rm b} h^2$ is ``physical'', as was already
noticed in \cite{Trotta:2003xg}. The redshift of decoupling $z_*$, for
instance, crucially depends on the number density of free electrons
$n_e$. Between helium and hydrogen recombination, this density is
related to the baryon number density $n_{\rm b}( \propto \omega_{\rm
  b})$ and Helium abundance by $n_e = n_{\rm b}(1 - Y_p)$. So a shift
in $z_*$ due to a change in $Y_p$ can be reversed by changing the
baryon density. The dependence of $n_e$ on the Helium fraction also
plays a r{\^o}le for the diffusion damping scale $d$, with $d \propto
n_e^{-1/2}$. An increase in $Y_p$, for example, means smaller $n_e$
and hence a larger $d$, which leads to an additional suppression of
power in the CMB temperature anisotropies on small
scales. Phenomenologically, this signature is similar to tilting the
spectrum of primordial anisotropies (i.e., lowering $n_{\rm S}$),
which explains the degeneracy with the spectral index.

The degeneracies with the Hubble parameter and the normalisation
are only indirect ones, since these parameters are themselves
correlated with the baryon density and the spectral index,
respectively. Note that the degeneracy with $A_{\rm S}$ is not
invariant under a change of the pivot scale. Had we chosen the
pivot at a small scale instead of a large scale, one would expect
a positive correlation instead of an anticorrelation between the
two parameters.

A comparison between the run with free $Y_p$ and that with a BBN prior
shows that imposing standard BBN places an extremely strong constraint
on the Helium fraction if we do not vary $\Delta N$. The width of the
68\% credible interval on $Y_p$ is $6.2 \times 10^{-4}$, i.e., the
error is dominated by the theoretical uncertainty in the prediction of
$Y_p^{\rm BBN}$ (see~table~\ref{table:sbbnerrors}).  Essentially, the
precise determination of the baryon density from {\sc planck} data
will also nail down the Helium fraction. This should not be
surprising, given that $Y_p^{\rm BBN}$ is relatively flat in the
direction of $\omega_{\rm b}$. Consequently, the expected errors on
the cosmological parameters hardly differ from the results of an
analysis with fixed $Y_p$, while the errors on the spectral index, the
normalisation, the Hubble parameter and the baryon density are up to a
factor two smaller than in the case with $Y_p$ as a free parameter.

\subsubsection{Extended model}
\begin{figure}
\begin{center}
    \includegraphics[height=\textwidth,angle=270]{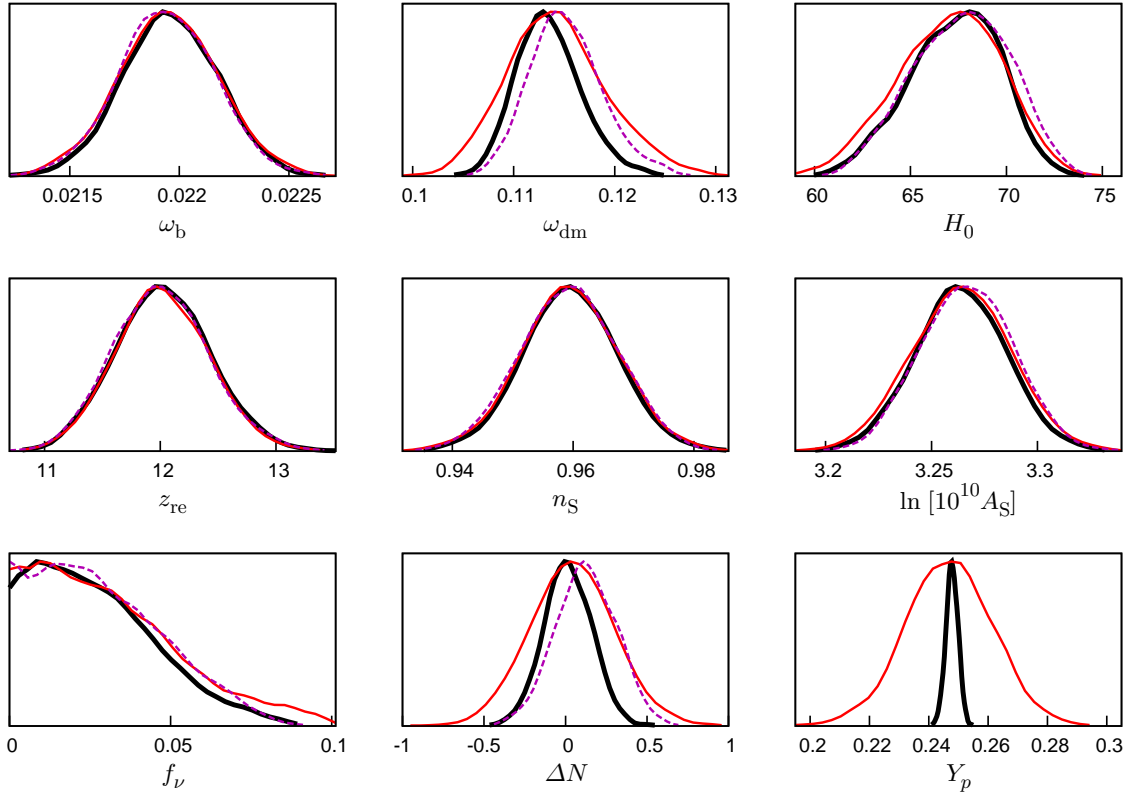}
\end{center}
\caption{\label{fig:9p1d} Marginalised posterior probabilities for
the parameters of the extended model, still with zero neutrino
chemical potential. The dashed purple curves correspond to the case
where $Y_p$ is fixed to a value of 0.24, the red curves have $Y_p$ as
a free parameter and the thick black curves correspond to the case
with a standard BBN prior.}
\end{figure}
Allowing $\Delta N$ to vary in the extended model
slightly weakens the constraining power of the standard BBN
consistency relation. Since $Y_p^{\rm BBN}$ is somewhat steeper in
the direction of $\Delta N$, the Helium fraction will be
allowed to vary over a wider range of values. Still, as we can see
from figure \ref{fig:9p1d}, the error on $Y_p$ is still
significantly improved (the width of the 68\% credible interval is
$4.1 \times 10^{-3}$ if we impose standard BBN, while for a free
$Y_p$ it is larger by a factor of seven).

By adding $f_\nu$ and $\Delta N$ to the free parameters of the model,
we naturally introduce new correlations which weaken the bounds on the
other parameters. The new degeneracies (particularly those with
$\Delta N$) turn out to be more serious than the degeneracies with the
Helium fraction. In models with free $\Delta N$, constraints on the
other cosmological parameters (apart from $\omega_{\rm dm}=\Omega_{\rm
dm} h^2$ and $\Delta N$ itself) are essentially independent on how
well the Helium fraction is constrained. Even fixing $Y_p$ to the
``wrong'' value of 0.24 does not lead to a bias worth mentioning. The
only exceptions are $\Delta N$ and the dark matter density (the latter
due to a correlation with $\Delta N$, see, e.g.~\cite{Hamann:2007pi}),
for which the error improves between ``free $Y_p$'' and ``BBN prior on
$Y_p$'' by a factor of 1.7 and 1.5, respectively.

Note that in these runs, not all posteriors peak at their fiducial
values. At first glance, this may sound strange since we use the
fiducial spectra in place of mock data spectra. However, the mismatch
just reflects the difference between the one-dimensional likelihood
profile and the marginalised posterior when the posterior is far from
gaussian in certain directions (like, in our case, the neutrino
fraction).

\subsection{Degenerate BBN \label{sec:dbbn}}

The effect of neutrino chemical potentials on BBN predictions is
twofold. It contributes to the radiation energy density so the
value of $N_{\rm eff}$ for three neutrinos with a common value of
$\xi$ becomes
\begin{equation}
N_{\rm eff} = 3.046 + 3 \left( \frac{30}{7} \left( \frac{\xi}{\pi}
\right)^2 + \frac{15}{7} \left( \frac{\xi}{\pi} \right)^4 \right)~,
\label{neff1}
\end{equation}
implying a larger expansion rate of the Universe, an earlier weak
process freeze out with a higher value for the neutron to proton
density ratio, and thus a larger value of $Y_p$. Subdominant
effects on neutrino decoupling \cite{Freese:1982ci,KangSteigman}
and their eventual distribution after the $e^+$-$e^-$ annihilation phase
\cite{Esposito:2000hi} are indeed very small and can be neglected.
On the other hand, a positive $\xi$ for the electron neutrino, i.e, a
larger number of $\nu_e$ with respect to $\bar{\nu}_e$, enhances $n
\rightarrow p$ weak processes compared to the inverse processes,
lowering the number of neutrons per proton available at the onset of
BBN.

Bounds on the value of the neutrino chemical potential from BBN data
analyses have been considered by many authors, see
e.g.~\cite{Freese:1982ci,KangSteigman,Esposito:2000hi,Esposito:2000hh,Hansen:2001hi,Barger:2003rt,Cuoco:2003cu,Serpico:2005bc}.
Since flavour oscillations enforce the condition of equal chemical
potentials for the three neutrino species \cite{Dolgov:2002ab}, the
(common) value of $\xi$ is strongly bounded by the neutron-proton beta
equilibrium and the observed value of $^4$He mass fraction. Adopting a
conservative error analysis of primordial $Y_p$ as in \cite{olive},
one gets $-0.04 \leq \xi \leq 0.07$ \cite{Serpico:2005bc}. This bound
can be evaded in non-standard scenarios with extra relativistic
degrees of freedom contributing to $\Delta N$, which is strongly
degenerate with $\xi$.

We now consider the effect of a dBBN prior on $Y_p$ in CMB data analysis,
in order to see what is the impact of the extra parameter $\xi$ on
future {\sc planck} estimates of cosmological parameters, including of
course the value of $\xi$ itself. As for the previous standard BBN
case, we do not use any direct experimental information on $Y_p$.

\subsubsection{Minimal model plus $\xi$}

We repeat our analysis for the minimal model with one extra free
parameter $\xi$, leaving $Y_p$ either free, or imposing a dBBN prior
on it. The constraint on cosmological parameters is essentially the
same in the case ``free $\xi \neq 0$ and BBN prior'' as in the case
``$\xi=0$ and no BBN prior''. In other words, by assuming dBBN instead
of standard BBN, the constraining power of the BBN prior on the six
basic $\Lambda$CDM cosmological parameters disappears.

\begin{table}[t]
\caption{\label{table:dbbnerrors} This table shows the same
quantities as in table \ref{table:sbbnerrors} for the degenerate
BBN scenario (with one extra parameter $\xi \equiv \mu_{\nu}/ k
T_{\nu}$). Note that $f_\nu = 0$ does not lie within the 68\% credible
interval for the extended model with free $Y_p$, so we quote the
limits of the interval instead of an upper bound.}  \vskip5mm \hskip35mm
\footnotesize{\begin{tabular}{lllll} \br &
\multicolumn{2}{c}{Minimal model} & \multicolumn{2}{c}{Extended
model}\\ & \multicolumn{1}{c}{free} & \multicolumn{1}{c}{BBN} &
\multicolumn{1}{c}{free} & \multicolumn{1}{c}{BBN}\\ \mr
$\Omega_{\rm b} h^2$ & $2.1 \times 10^{-4}$ & $2.1 \times 10^{-4}$
& $2.2 \times 10^{-4}$ & $2.3 \times 10^{-4}$\\ 
$\Omega_{\rm dm} h^2$ & $2.1 \times 10^{-3}$ & $1.3 \times 10^{-3}$ &
$5.0 \times  10^{-3}$ & $4.9 \times 10^{-3}$ \\ 
$h$ & $10. \times 10^{-3}$ & $7.3 \times 10^{-3}$ & $28 \times
10^{-3}$ & $27 \times 10^{-3}$\\  
$z_\mathrm{re}$ & 0.41 & 0.40 & 0.43 & 0.41 \\
$\ln[10^{10}A_\mathrm{S}]$ & 0.023 & 0.022 & 0.023 & 0.024\\
$n_\mathrm{S}$ & $7.4 \times 10^{-3}$ & $6.9 \times 10^{-3}$ &
$8.0 \times 10^{-3}$ & $8.1 \times 10^{-3}$ \\ 
$Y_p$ & 0.012 & 0.010 & 0.016 & 0.016 \\ 
$\xi$ & 0.34 & 0.061 & 0.45 & 0.093 \\
$\Delta N$ & -- & -- & 0.27 & 0.27 \\ 
$f_\nu$ & -- & -- & $ 0.016 \to 0.063$ & $\leq 0.039$ \\ \br
\end{tabular}}
\end{table}

However, it is interesting to see how the degenerate BBN prior
improves the constraint on $\xi$ itself. To this end one can compare
the results of the ``free'' and ``BBN'' cases of table
\ref{table:dbbnerrors} (both for the minimal model and the extended
one). The projected error on $Y_p$ does not change noticeably between
the two cases, since now the $^4$He mass fraction is not only
determined by the baryon density and $\Delta N$, but also depends on
the value of $\xi$. Even very small departure from zero allows quite a
large variation of $Y_p$, therefore there is not much difference in
this case between imposing the BBN relation and leaving the Helium
mass fraction a free parameter. In turn, this strong dependence of
$Y_p$ on $\xi$ is also responsible for the significant decrease of the
error on $\xi$ in the BBN case. Indeed, the CMB alone is able to
constrain $\xi$ only through its contribution to the total energy
density during radiation domination, see equation~(\ref{neff1}). In
this case the 68\% error is $\sigma(\xi)=0.34$, consistent with
previous analyses based on the Fisher matrix
approximation~\cite{Kinney:1999pd,Lesgourgues:1999ej,Bowen:2001in}.
When including a BBN prior, the sensitivity of the CMB to $Y_p$
further reduces the error down to $\sigma(\xi)=0.061$. Remarkably,
this result is comparable to the error expected using observations of
light element abundances, in case one adopts a conservative approach
on the $Y_p$ error estimate to account for possible systematics. We
conclude that future CMB data like that from {\sc planck} will be a
very useful probe of the neutrino asymmetry. In case of a $\xi
\neq 0$ detection, it would be of particular interest to compare this
finding with constraints from primordial element data.

\subsubsection{Extended model plus $\xi$}

\begin{figure}
\begin{center}
    \includegraphics[height=0.7\textwidth,angle=270]{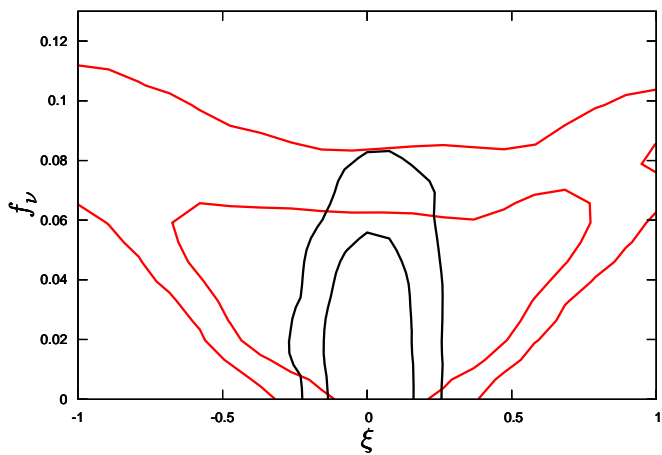}
\end{center}
\caption{\label{fig:xifnu} Two-dimensional marginalised joint
posterior 68\%- and 95\%-credible contours for $\xi$ and $f_\nu$ in
the extended model with free neutrino chemical potential. Red lines
correspond to ``free $Y_p$", the thin black lines are the results of
imposing the BBN prior.}
\end{figure}

Lastly, we analyse the extended model with a neutrino chemical
potential. This requires some particular modifications of
\texttt{CAMB} in order to explicitly include the chemical potential of
neutrinos and anti-neutrinos in the expression of the massive neutrino
phase-space distribution (as explained in \cite{Lesgourgues:1999wu}).
The total density during radiation domination is now parameterised as
\begin{equation}
N_{\rm eff} = 3 \, \left( 1 + \frac{30}{7} \left( \frac{\xi}{\pi}
\right)^2 + \frac{15}{7} \left( \frac{\xi}{\pi} \right)^4 \right)
+ \Delta N~.
\end{equation}
Our results are shown in table \ref{table:dbbnerrors}. Note that due
to slow convergence of the chains for these models, we relaxed
our convergence criterion slightly, demanding $R-1 \leq 0.07$. As for
the minimal model, the value of $Y_p$ is determined with the same
(poor) accuracy regardless of whether one imposes the dBBN prior or
leaves the Helium mass fraction as a free parameter, while the error
on $\xi$ is strongly reduced in the first case, because of the strong
dependence of $Y_p$ on this parameter. The only other parameter
affected is the neutrino mass fraction $f_\nu$. Its degeneracy with
$\xi$ is illustrated in figure~\ref{fig:xifnu}. In fact, if one were
to include large scale structure data the bounds on $\xi$ would likely
be significantly further reduced, because the degeneracy existing
between $f_{\nu}$ and $N_{\rm eff}$ would be alleviated. Notice also
that the error on $\Delta N$ is basically unchanged, as the dependence
of $Y_p$ on this parameter is much weaker.
Finally, it is also worth comparing the results for the extended
models when imposing the standard or degenerate BBN priors, i.e.
the last column of tables \ref{table:sbbnerrors} and
\ref{table:dbbnerrors}. With the exception of the dark matter density,
the basic $\Lambda$CDM cosmological parameter errors do not undergo
any substantial change, while we again see the effect of the
introduction of the extra parameter $\xi$ in reducing the sensitivity
to $Y_p$.  Correspondingly, the error on $\Delta N$, and, due to the
aforementioned correlation, also the error on $\Omega_{\rm dm}h^2$,
grow by almost a factor of two with respect to the standard BBN
scenario, reaching the values which are obtained if $Y_p$ is taken as
a free parameter and $\xi=0$, see table \ref{table:sbbnerrors}.

\section{Conclusions}\label{sec:conclusions}

In this paper we have considered in detail the effects of a careful
calculation of the Helium mass fraction on future CMB experiments,
such as {\sc planck}, using the known fact that $Y_p$ is not an
independent free parameter, but can rather be fixed in the framework
of Big Bang Nucleosynthesis as a function of the baryon density, the
energy density during the relativistic dominated era, as well as, in
more exotic scenarios, other physical inputs such as neutrino
degeneracy. In view of the high precision in parameter estimates which
is expected to be achieved by {\sc planck}, this method first adopted
in \cite{Ichikawa:2006dt} is more consistent than the current strategy
of fixing by hand the value of $Y_p$. It is also worth stressing that
it is completely independent from any astrophysical information on
light nuclei abundance, as $Y_p$ determination from low metallicity
HII regions in blue compact galaxies. In this respect, this method is
different from combined analyses of CMB and BBN data (as performed for
example in
\cite{Hansen:2001hi,Barger:2003zg,Cuoco:2003cu,Cyburt:2004yc,Mangano:2006ur}).

We have considered two different models, the standard BBN
scenario, with possibly extra relativistic species in addition to
three standard active neutrinos, and the case of degenerate BBN,
with sizable neutrino chemical potentials.

For standard BBN, $Y_p$ only depends on the baryon fraction and
$\Delta N$. Exploiting this functional dependence in a forecast
for {\sc planck} data, we have shown that one can avoid a possible
bias in the estimate of some cosmological parameters, i.e. the
spectral index $n_S$ and $A_S$, which is instead present if $Y_p$
is fixed a priori to some reference value, usually given by
$Y_p=0.24$. Furthermore, this method allows for a better
determination of various parameters (like e.g. the baryon density,
the spectral index, the number of extra relativistic degrees of
freedom and of course $Y_p$ itself, see table
\ref{table:sbbnerrors}).  With a BBN prior, the Helium mass
fraction can be determined with an accuracy better than 1~$\%$, at
the level of statistical error of astrophysical determinations,
which are however possibly plagued by a larger systematic error.
On the other hand, without imposing the BBN prior, CMB data from
{\sc planck} can determine $Y_p$ at the 5-6~$\%$ level only.

In the case of degenerate BBN, due to the strong dependence of $Y_p$
on one extra parameter -- namely, the neutrino chemical potential
parameter $\xi$ assumed to be flavour independent due to flavour
oscillations -- this result is no longer valid. Imposing the BBN
prior, the value of $Y_p$ is only determined with an order 10~$\%$
uncertainty, just as if a flat prior was assumed over the whole range
$0 \leq Y_p \leq 1$. Nevertheless, exploiting the dependence of the
Helium mass fraction on $\xi$ has a big impact on the way this
parameter can be determined by CMB anisotropy data. For a fiducial
value $\xi=0$, we found that the 68$\%$ absolute error on this
parameter is 0.06 for the minimal model with no extra radiation, and
0.09 for the extended model where $\Delta N$ and the neutrino mass
fraction $f_\nu$ are allowed to vary. In both cases, the result is
comparable with the error obtained when using nuclei abundance data
alone. If $Y_p$ is assumed to be an independent parameter, with no BBN
prior, the effect of $\xi$ on the CMB power spectrum is only via its
contribution to the relativistic energy density which shifts the
matter-radiation equivalence point. Actually, in this case the 68$\%$
bound is up to one order of magnitude weaker, $|\xi| \leq 0.45$, again
for our fiducial model with $\xi=0$.  Further spectroscopic
measurement of $^4$He abundance or a better understanding of
systematics effects would of course, provide a powerful way of
independently constraining (or measuring) the lepton asymmetry in the
neutrino sector. Yet comparison with future CMB data will represent an
important consistency check.

\section*{Acknowledgments}

We thank Yvonne Wong for providing us with her fiducial data
generation code. GM thanks Universit\'e de Savoie for a one
month visitorship at LAPTH during which most of this project was
completed. JL acknowledges support from the EU 6th
Framework Marie Curie Research and Training network ``UniverseNet''
(MRTN-CT-2006-035863). Numerical simulations were performed on the
MUST cluster at LAPP (IN2P3/CNRS and Universit\'e de Savoie). JH was
supported by the ANR (Agence Nationale de la Recherche).

\section*{References}

\end{document}